\documentclass[english]{article}
\usepackage[utf8]{inputenc}
\usepackage[T1]{fontenc}
\usepackage{babel}
\usepackage{amsmath}
\usepackage{graphicx}
\usepackage{fancyhdr}
\usepackage{pifont}
\usepackage{float}
\usepackage{rotating}
\PassOptionsToPackage{hyphens}{url}\usepackage{hyperref}
\usepackage{pmboxdraw}
\usepackage{tabularx}
\usepackage[dvipsnames]{xcolor}

\newcommand{\cmark}{\ding{51}}
\newcommand{\xmark}{\ding{55}}

\begin{document}

\title{BIMCV COVID-19+: a large annotated dataset of RX and CT images from COVID-19 patients} 

\author{Maria de la Iglesia Vay\'{a}\textsuperscript{1{*}},
Jose Manuel Saborit\textsuperscript{1},\\
Joaquim Angel Montell\textsuperscript{1},
Antonio Pertusa\textsuperscript{2{*}}, 
Aurelia Bustos\textsuperscript{3{*}}, \\
Miguel Cazorla\textsuperscript{2}, 
Joaquin Galant\textsuperscript{4}, Xavier Barber\textsuperscript{5}, \\
Domingo Orozco-Beltr\'{a}n\textsuperscript{5{*}}, 
Francisco García-García\textsuperscript{1,7},\\ Marisa Caparr\'{o}s\textsuperscript{1}, 
Germ\'{a}n Gonz\'alez\textsuperscript{2,6{*}}, Jose Mar\'ia Salinas\textsuperscript{1,4{*}}
}

\maketitle
\thispagestyle{fancy}
1. Unidad Mixta de Imagen Biomédica FISABIO-CIPF. Fundación para el Fomento de la Investigación Sanitario y Biomédica de la Comunidad Valenciana. Valencia, Spain.
2. Universidad de  Alicante, Spain.
3. Medbravo.
4. Hospital San Juan de Alicante, Spain.
5. Universidad Miguel Hern\'{a}ndez, Spain.
6. Sierra Research SL.
7. Bioinformatics \& Biostatistics Unit, Principe Felipe Research Center, Valencia, Spain. {*}corresponding authors: Maria de la Iglesia \\ (delaiglesia\_mar@gva.es), Antonio Pertusa (pertusa@ua.es), Aurelia Bustos (aurelia@medbravo.org),
Domingo Orozco-Beltr\'{a}n (dorozco@umh.es),  Germ\'{a}n Gonzalez (ggonzale@sierra-research.com), Jose Maria Salinas (salinas\_josser@gva.es)

\begin{abstract}

This paper describes BIMCV COVID-19+, a large dataset from the Valencian Region Medical ImageBank (BIMCV) containing chest X-ray images CXR (CR, DX) and computed tomography (CT) imaging of COVID-19+ patients along with their radiological findings and locations, pathologies, radiological reports (in Spanish), DICOM metadata, Polymerase chain reaction (PCR), Immunoglobulin G (IgG) and Immunoglobulin M (IgM) diagnostic antibody tests.  The findings have been mapped onto standard Unified Medical Language System (UMLS) terminology and cover a wide spectrum of thoracic entities, unlike the considerably more reduced number of entities annotated in previous datasets. Images are stored in high resolution and entities are localized with anatomical labels and stored in a Medical Imaging Data Structure (MIDS) format. In addition, 10 images were annotated by a team of radiologists to include semantic segmentation of radiological findings. 
This first iteration of the database includes 1,380 CX, 885 DX and 163 CT studies from 1,311 COVID-19+ patients.
This is, to the best of our knowledge, the largest COVID-19+ dataset of images available in an open format. The dataset can be downloaded from \url{http://bimcv.cipf.es/bimcv-projects/bimcv-covid19/}. 

\end{abstract}

\begin{table}[hbp]
    \centering
    \begin{tabularx}{\textwidth}{|X|X|}
        \hline
        \includegraphics[width=0.45\textwidth]{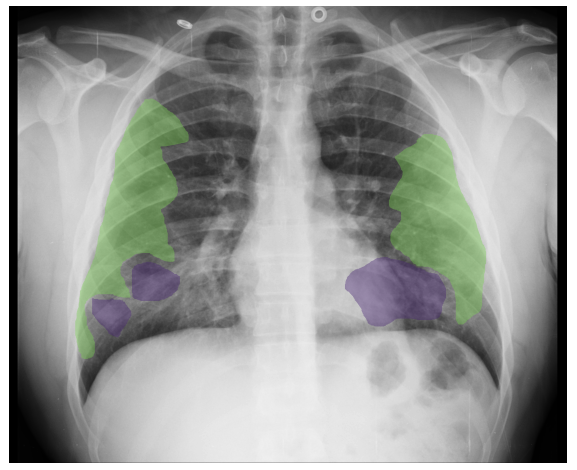}  & 
        
        \vspace{-4.25cm} 

        \textbf{Report:} opacidades de aspecto intersticioalveolar parcheadas y bilaterales que predominan en ambos lobulos inferiores  sospechosas de infeccion por COVID-19 .  senos costofrenicos libres .
        
        \textbf{Labels:} COVID 19, alveolar pattern, interstitial pattern, pneumonia
        
        \textbf{Locations:} costophrenic angle, lobar, bilateral, lower lobe

        \\ \hline
        \begin{tabular}{l}
            DICOM Fields\\
            \hline
            Study Date                                     20200317 \\
            Patient's Sex                                         M \\
            Patient's Birth Date                           1986 \\
            Modality                                             CR \\
            Manufacturer                          GE Healthcare \\
            ...
        \end{tabular}
        & 
        \begin{tabular}{lll}
Date & Test & Result \\
\hline
17.03.2020 &    PCR &  NEGATIVE \\
18.03.2020 &    PCR &  NEGATIVE \\
19.03.2020 &    IGG &  POSITIVE \\
19.03.2020 &    IGM &  POSITIVE \\
20.03.2020 &    PCR &  POSITIVE \\
& $\cdots$ & $\cdots$ \\
\hline
\end{tabular}
                \\ \hline
    \end{tabularx}
    \caption{Example of an image from the dataset and its associated information. Top left: image showing the Regions of Interest (ROIs) regarding ground glass opacities (green) and consolidations (purple) as marked by a trained radiologist. Top right: radiology report (in Spanish), radiological findings, differential diagnosis and locations as extracted using natural language processing from the radiology report. Bottom left: example of the image metadata as obtained from the DICOM fields. Bottom right: diagnostic tests performed on this subject, showing positive results for IGG and IGM on 19/03/2020 and positive to PCR on 20/03/2020. }
\end{table}



\section*{Background \& Summary}


SARS-Cov-2 has created an unprecedented pandemic situation. The scientific community has focused on the development of artificial intelligence (AI) algorithms for the better diagnosis and prognosis of COVID-19+ patients, but these efforts are often performed on proprietary datasets, whereas few image studies of patients positive to COVID-19 are publicly available. 

During the first half of March 2020, the American College of Radiology (ACR) published guidelines concerning medical imaging for COVID-19 diagnosis \cite{ACR}. Some of its recommendations include the following: ``CT should not be used to screen for or as a first-line test to diagnose COVID-19, CT should be used sparingly and reserved for hospitalized, symptomatic patients with specific clinical indications for CT''. This recommendation coincides with our initial commitment to focus on conventional radiology as a tool to aid the diagnosis, prognosis and triage of COVID-19 patients.

The BIMCV COVID-19+ dataset is a large open multi-institutional data-bank that provides the open scientific community with data of clinical-scientific value that will help the early detection and evolution of COVID-19. Given the current COVID-19 pandemic, both speed and efficiency in developing accurate medical solutions are key factors that will enable them to reach the clinical environment in the shortest possible time. Making the information accessible to the scientific community worldwide will undoubtedly maximize the usefulness of the data. Some well-known examples of the scope and benefit of open access medical data-sets are Pan-cancer Atlas (TCGA) \cite{liu2018integrated} of the National Cancer Institute, the MIMIC data-set \cite{beaulieu2018mapping, johnson2019mimic} by the MIT Lab for Computational Physiology, and Padchest \cite{padchest} from BIMCV and the University of Alicante.

The proposed dataset is intended to be incremental, and new images along with their annotations will, therefore, be continuously added when available. At this time (first iteration), BIMCV COVID-19+ contains 1,380 CR, 885 DX, and 163 CT full-resolution images from 1,311 patients. 

The data provided for each study, detailed in Sec. Data Records, include the images, anonymized DICOM metadata, anonymized radiologic reports (in Spanish) and UMLS biomedical vocabulary unique identifiers (CUIs) associated to each image (for example, `infiltrate', 'pleural effusion', etc.) organized in semantic trees.  In addition, a team of radiologists has manually annotated 10 images  with the Regions of Interest (ROI) of the findings that are related to COVID-19.

Regarding previous COVID-19 datasets, Table \ref{table:related_work_dataset} summarizes the main features of the datasets published to date. First, we describe the public datasets.

\begin{table*}[!htb]
\centering
\caption{Main characteristics of the different datasets containing COVID-19 information published to date. The highlighted features are: \#Imag. number of COVID-19 images; \#Pat. number of patients; Sex; Age; Diagnostics: COVID, No (COVID), Others (pneumonia caused by other virus); Surv. information concerning the patient survival; View: Patient position; \#Rx: number of Rx images; \#CT: number of CT images;  Rad.Rep.: Radiological Reports; Pub.: Public.\label{table:related_work_dataset}}

\begin{small}
\begin{tabular}{lccccc}
\hline
Dataset & \#Imag. & \#Pat. & Sex & Age & Diagnostics \\
\hline
COVID-CHESTXRAY \cite{cohen2020covid19} & 373 & 205 & \cmark & \cmark & COVID/Others\\
COVID-CT \cite{zhao2020COVID-CT-Dataset} & 349 & 216 & \cmark & \cmark & COVID/No \\
SIRM-COVID \cite{SIRM} &  340 & 85 & \cmark & \cmark & COVID \\
COVID-19 RAD. DB.  \cite{chowdhury2020ai} & 219 & Unknown & \xmark &\xmark & COVID/No/Others \\
Private dataset \cite{li2020artificial} & 1,296 & Unknown &\xmark &\xmark & COVID/No/Others\\
Private dataset \cite{jin2020ai} & 877 & Unknown &\xmark &\xmark & COVID/No \\
Private dataset \cite{shi2020large} & 1,658 & Unknown &\xmark &\xmark & COVID/No/Others \\

\textbf{BIMCV COVID-19+} & \textbf{5,381} & \textbf{1,311} & \cmark & \cmark & COVID/No/Others\\
\hline

\end{tabular}

\vspace{0.5cm}
\begin{tabular}{lcccccc}
\hline
Dataset &  Surv. & View & Rx & CT &  Rad.Rep. & Pub. \\
\hline
COVID-CHESTXRAY \cite{cohen2020covid19} & \cmark &  \cmark & 10 &10 & \cmark&  \cmark \\
COVID-CT \cite{zhao2020COVID-CT-Dataset} &  \xmark &\xmark & \xmark & 349 &  \cmark&  \cmark \\
SIRM-COVID \cite{SIRM} &  \xmark &\cmark & 255 & 85 &  \cmark &  \cmark \\
COVID-19 RAD. DB. \cite{chowdhury2020ai} & \xmark &\xmark & 219 & \xmark &\xmark & \cmark\\
Private dataset \cite{li2020artificial} & \xmark &\xmark &\xmark &1296 &\xmark &\xmark \\
Private dataset \cite{jin2020ai} & \xmark &\xmark & \xmark & 877 & \xmark & \xmark\\
Private dataset \cite{shi2020large} & \xmark &\xmark & \xmark & 1658 & \xmark & \xmark\\
\textbf{BIMCV COVID-19+} & \xmark & \cmark & \textbf{3,141} & \textbf{2,239}  & \cmark  &  \cmark  \\
\hline
\end{tabular}
\end{small}
\end{table*}

COVID-CHESTXRAY \cite{cohen2020covid19} is a public dataset of pneumonia cases with chest X-ray or CT images, specifically COVID-19 cases, along with MERS, SARS, and ARDS. Data are  collected from public sources in order not to infringe patient confidentiality, and it contains data from 205 patients. 
COVID-CT \cite{zhao2020COVID-CT-Dataset} is a CT public dataset. Images are collected from COVID-19 related publications, and are extracted directly from papers in PDF format. The quality of images is not, therefore, optimal. The number of patients is similar that stated above.
SIRM-COVID \cite{SIRM} is a dataset of COVID-19 cases with RX and CT images. It contains only positive cases, and includes both radiological and clinical reports. COVID-19 RADIOGRAPHY DATABASE \cite{chowdhury2020ai} is published in Kaggle and  contains a large number of normal Rx (1,341) and other viral pneumonia images (1,345). However, it contains few COVID-19 positive images.

There are also some private datasets. 
The authors of \cite{li2020artificial} describe a non-public dataset that contains a large number of CT images, along with 1,735 samples for community acquired pneumonia and 1,325 for non-pneumonia. Another private CT dataset is described in \cite{jin2020ai}. In this case, it contains 877 COVID-19+ images and 541 COVID-. Finally, \cite{shi2020large} also has a private CT dataset, with 1,659 positives and 1,027 with community acquired pneumonia.

In summary, to the best of our knowledge, BIMCV COVID-19+ is the first dataset that includes radiological findings. In addition, it is the only COVID-19 bank in which some images are  labeled ROIs annotated by radiologists. Moreover, it is the largest dataset as regards number of images and patients, and includes multiple samples for each patient in order to analyze their clinical evolution. The dataset is intended to be incremental and it will grow as soon as images are  available. This work describes the dataset at its first iteration (version 1), with release date 28/05/2020.

\section*{Methods}

This section addresses the methodology used to produce the data. We also describe the ethics involved in the process of data acquisition, the data anonymization, and the labeling process.

\subsection*{Ethics statement}
The Institutional Review Board (IRB) of the Miguel Hernandez University (MHU) approved this HIPAA-compliant retrospective cohort study. The study was approved by the local institutional ethics committee CElm: 12/2020 at Arnau de Vilanova Hospital in Valencia Region. The healthcare authorities of the Comunitat Valenciana authorized the publication of the open database based on the basis of different reports that had to be written, of which the following should be highlighted: a favorable report from the Data Protection Officer (DPO); a development report of an Impact study; a PIs confidentiality agreement; a confidentiality agreement between researchers; the declaration of a data protection officer for all participating entities; a compliance report of security measures proposed in the DPO report and, finally, a detailed risk study together with mitigation measures for detected risks.

\subsection*{Data acquisition}

The data was acquired following the process described in \cite{CEIB}. R \& D Cloud CEIB has four general modules: Search engine (SE), manager of clinical trials (GEBID), anonymizer (CEIBANON) and BioImage Knowledge Engine (BIKE). The technology used in R \& D Cloud CEIB is based completely on Open Source. 

All consecutive studies of patients with at least one positive PCR test or positive immunological tests (IgM, IgG or IgA) for SARS-Cov-2 in the period of time between February 26th and April 18th, 2020 were identified by querying the Laboratory Information System records from the Health Information Systems in the Comunitat Valenciana. All medical images acquired for such subjects in this period of time were included in the BIMCV COVID-19+ dataset. The images included chest x rays (both digital DX and digitalized film CR), along with computed tomography (CT) scans. The images were retrieved from the Vendor Neutral Archive (VNA) in the central medical image repository of the Digital Medical Image Management project (GIMD) appertaining to the public healthcare system in the Valencian Region (Spain). The data in its first iteration were obtained from 11 hospitals from the Valencian Region.

\subsection*{Data Geo-positioning}

The Valencian healthcare system is arranged in health departments, which are equivalent to the health areas provided in the General Health Law. The health departments are the fundamental structures of the Valencian healthcare system, being the geographical demarcations into which the territory of the Valencian Region is divided for healthcare purposes and constituting the framework for the integration of healthcare promotion and protection actions, prevention and cure and rehabilitation of health status, through the coordination of existing resources and guaranteeing healthcare. 

\begin{figure}[!htb]
    \centering
    \includegraphics[width=0.6\columnwidth]{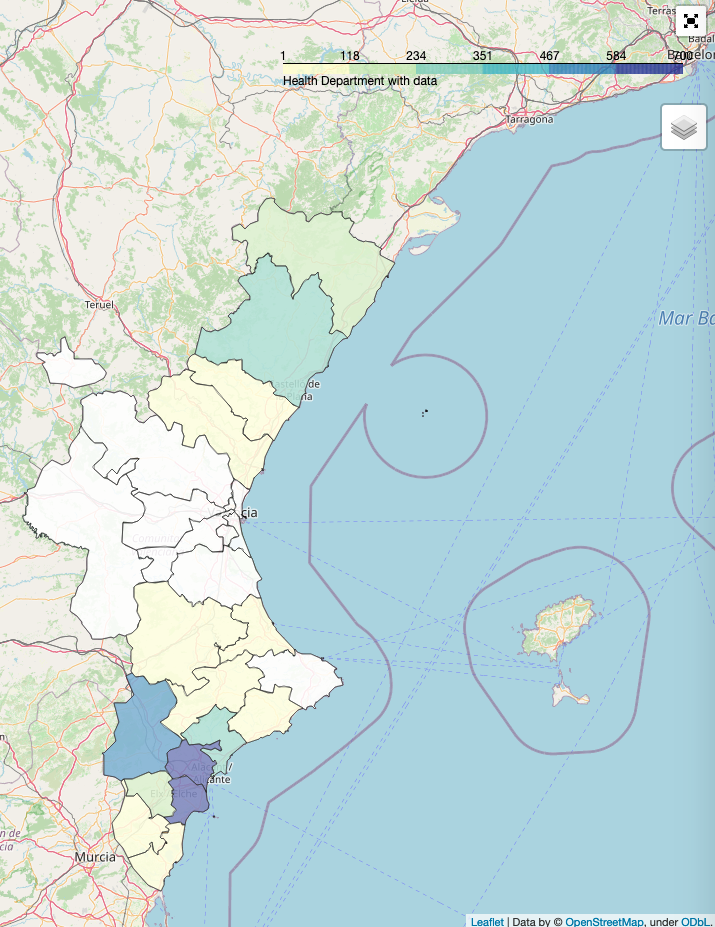}
    \caption{Choropletic map with \#CR \#DX \#CT from first iteration. \url{https://maigva.github.io/maps/HealthDepartCOVID19.html}}
    \label{fig:geopos}
\end{figure}

Figure \ref{fig:geopos} shows the choropleth map containing the number of tests performed (CR, DX and CT) by the health departments. In this first iteration of the dataset, it will be noted that most of the departments (11 concretely) are in the provinces of Alicante and Castellón. For future iterations, we plan to complete the map.

\subsection*{Data anonymization}

The Organic Law 3/2018 \cite{BOE} establishes the legal framework for data protection in biomedical research. The reuse of personal data for medical research must be approved by an ethics committee, and data must be at least pseudonymized before the researchers attain access. In this respect, anonymization is defined as ``the result of the processing of personal data to irreversibly prevent its identification'', which is why anonymization in itself constitutes the additional processing of personal data. Within this complex and sensitive framework, state-of-the-art anonymization techniques in Medical Imaging have been applied to the BIMCV COVID-19+ dataset. The data were anonymized in two ways. First, all references to patient data were removed from the radiological report by using the methodology explained in \cite{Perez-Diez2020.04.09.20058958}. This method is designed to parse Spanish reports. It uses a pre-trained Bidirectional LSTM to identify names and personal data and remove them from the report.

Second, DICOM anonymization was performed based on DICOM PS3.15 Annex E using a CTP \cite{aryanto2012implementation} server. Finally, the images were evaluated in a semi-automatic process and some of them were visually inspected to remove or crop the burnt-in personal information from the chest x-ray images.

\subsubsection*{Radiological reports anonymization}
\label{subsub:radio_anon}

Anonymization of radiological reports was performed by applying the DisMed methodology based on a Named Entity Recognition (NER) strategy, which is focused on the extraction and location of seven predefined entities or categories (see Table \ref{tab:anonymentities}) belonging to radiological reports and with a set-up of this system to be adapted to the domain of medical texts.

\begin{figure}[!htb]
    \centering
    \includegraphics[width=\columnwidth]{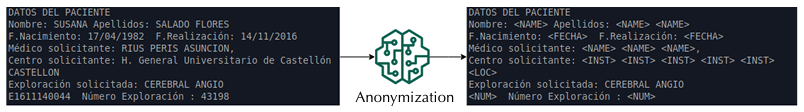}
    \caption{Conceptual scheme of radiological report anonymization.}
    \label{fig:anonym}
\end{figure}

\begin{table}[!htb]
\caption{\label{tab:anonymentities}  Name Entities selected for anonymization and their associated Protected Health Information categories.}
\begin{tabular}{l|p{6cm}|p{4cm}}
NEs   & Description                                                                                                                                                     & PHIs                                                                                                    \\
\hline
CAB   & Section headers                                                                                                                                                 &                                                                                                         \\
NAME  & Names and surnames (patient and others)                                                                                                                         & Names                                                                                                   \\
DIR   & Full addresses, including streets, numbers and zip codes                                                                                                        & Geographic data                                                                                         \\
LOC   & Cities, inside and outside addresses                                                                                                                            & Geographic data                                                                                         \\
NUM   & Numbers or alphanumeric strings that might identify someone,  including digital signatures, patient numbers, medical numbers, medical  license numbers and others & Medical record numbers, social security numbers, account numbers, any unique identifying number or code \\
FECHA & Dates                                                                       & Dates                                                                                                   \\
INST  & Hospitals, health centers or other institutions                                                                                                                 &           
\end{tabular}
\end{table}

\subsubsection*{DICOM anonymization}

The DICOM standard has already defined the framework that guarantees the safe access, exchange and processing of medical data. However, to the best of our knowledge, there are no tools that implement all the confidentiality profiles defined in Part 15 of the DICOM standard \cite{aryanto2012implementation, shi2018development} in paragraph E 'Attribute Confidentiality Profiles', with the exception of the Clinical Trial Processor (CTP \cite{10.1145/3239438.3239449}).
To avoid compromising data privacy, it is considered crucial to implement different modules and scripts within the CTP that will comprise the development of the ten DICOM confidentiality profiles that are defined in the standard.



The scripts related to the Basic Profile have been applied to each module. This is the strictest profile and eliminates all the information related to (i) the identity, along with the identifying and demographic characteristics of the patient; (ii) the identity of the authors, person responsible or relatives; (iii) the identity of any personnel involved in the procedure; (iv) the identity of the organizations involved in ordering or carrying out the method; (v) the information (not related to the patient) that could be used to discover the identity of the original anonymized files (for example, UID, date and time), and (vi) private attributes (which are not part of the standard).

The system permits different types of anonymity, from the alteration of the existing text information in DICOM headers up to image-level deformation of parts that can identify the patient (especially in neuroimaging obtained by magnetic resonance). In the case of RX/DX/CT, no image deformations were necessary since it is not possible to reconstruct patients' data from this information.

\subsection*{Image preprocessing}
\label{subsub:preprocessing}
Raw pixel data were extracted from the DICOM images and stored in files in \textit{nii.gz} format. The images were processed by rescaling the dynamic range using the DICOM window width and center, when available, and stored as 16-bit PNG images. The images were not re-scaled, to avoid loss of resolution. 

The information on image projection was estimated using a neural network with an EfficientNet \cite{tan2019efficientnet} architecture. The projection estimation network was trained with $2,000$ images from Padchest \cite{padchest}, for which the orientation was estimated manually. The projection was oversimplified to frontal (including antero-posterior and postero-anterior) and lateral (L). No differentiation was made between erect, either standing or sitting or decubitus. No differentiation was made between left lateral or right lateral. Such information could subsequently be inferred by a careful analysis of the DICOM tags.

\subsection*{Labeling}

The early detection and location of lesions such as infiltrates, and particularly ground glass opacities, is essential as regards diagnosis, predicting the evolution of the patient and helping make clinical decisions. The findings that suggest a COVID-19 infection are both the ground-glass opacity and consolidation that, even in the initial stages, affect both lungs, particularly the lower lobes, and especially the posterior segments, mainly with a peripheral and subpleural distribution  \cite{Bernheim20, Pan20, Wang20, Zhao20}. 


These findings were present on chest CT scans in 44\% of patients in the first two days  \cite{Bernheim20}, in 75\% of patients in the first four days \cite{Pan20} and in 86\% of patients during illness days 0-5 \cite{Wang20}.

These lesions can progress in the following days until they become more diffuse. If associated with septal thickening, they will have a reticular pattern. In general, they progress both in extension and towards consolidation, concomitantly with the ground glass pattern, and may then have a rounded morphology. It is very rare for them to be associated with lymphadenopathy, cavitation or pneumothorax, as occurred with the respiratory syndrome coronavirus MERS-CoV \cite{mmg20}.

Beyond labeling an image as COVID/No/Others as occurs with most existing datasets, it is, therefore, very important to annotate the radiological findings, for both the diagnosis and the prognosis. It should be considered that these characteristic lesions frequently appear in non-COVID patients and that, on the contrary, many COVID-19 patients do not have these lesions. The appearance of the lesions is related to the level of the disease. It is characteristic that many of the patients worsen and that lesions appear or become more evident. 

To the best of our knowledge, BIMCV COVID-19+ is the first COVID-19 dataset that includes these findings. In addition, in this first iteration, ten images were also annotated with the ROIs of the findings.

\subsubsection*{Assigning CUI labels to images}
\label{subsub:CUI}

Clinical reports were used as a basis to obtain the radiological findings, differential diagnoses and localization labels. Therefore, images were not used in this process. 

The labels were obtained using the same procedure employed to build PadChest \cite{padchest}, but by  adding to the set of labels the COVID-19 (CUI C5203670) and COVID-19 uncertain (CUI C5203671) to the set of labels, leading to a total of 336 different labels. After a preprocessing stage in which the words from the radiological report were stemmed and tokenized, labels were automatically extracted using a deep neural network multi-label classifier trained with the PadChest dataset. The network topology is detailed in \cite{padchest}, but was re-trained with the additional COVID-19 labels. 

More specifically, this classifier consists of a bidirectional LSTM with an attention mechanism that receives a preprocessed radiological report encoded as a sequence of pre-trained word-embeddings, and outputs the most likely labels. Labels corresponding to diagnoses, radiological findings and anatomical localizations are mapped onto the UMLS controlled biomedical vocabulary unique identifiers (CUIs) and organized into semantic concept trees.

For example, consider the following report sentence: \textit{Cambio pulmonar crónico severo. Signos de fibrosis bibasal. Sutil infiltrado pseudonodular milimétrico en vidrio deslustrado localizado en bases. Cifosis severa.}

In this case, the neural network would outputs the following labels: ['pulmonary fibrosis',  'chronic changes',  'kyphosis',  'pseudonodule',  'ground glass pattern'], along with their localizations when available: [['pulmonary fibrosis',  'loc basal bilateral'],  ['chronic changes'],  ['kyphosis'],  ['pseudonodule',  'ground glass pattern',  'loc basal']].

An example of a PadChest semantic concept sub-tree of some lesions that are frequently associated with COVID-19 is shown below:

\begin{footnotesize}
\noindent
├── infiltrates [CUI:C0277877]\\
│\verb!   !├── interstitial pattern [CUI:C2073538]\\
│\verb!   !│\verb!   !├── ground glass pattern [CUI:C3544344]\\
│\verb!   !│\verb!   !├── reticular interstitial pattern []\\
│\verb!   !│\verb!   !├── reticulonodular interstitial pattern [CUI:C2073672]\\
│\verb!   !│\verb!   !└── miliary opacities [CUI:C2073583]\\
│\verb!   !└── alveolar pattern [CUI:C1332240]\\
│\verb!   !\verb!   ! ├── consolidation [CUI:C0521530]\\
│\verb!   !\verb!   ! │\verb!   !└── air bronchogram [CUI:C3669021]\\
│\verb!   !\verb!   ! └── air bronchogram [CUI:C3669021]\\
\end{footnotesize}

As will be noted, although the radiological reports are provided in Spanish, labels are mapped onto biomedical vocabulary unique identifier (CUIs) codes, thus making the dataset usable regardless of the language.

In addition to the PadChest labels that can be found in \cite{padchest}, the new ``COVID-19'' and ``COVID-19 uncertain'' labels were added as new terms, to denote high or low suspicion of COVID-19 respectively. To this end, the word-embeddings of these new labels were pre-computed in the corpus of new reports for COVID-19 patients added to the Padchest Spanish report corpus. The Padchest multi-label text classifier was then trained with the manually labeled reports from Padchest, but adding 724 new manually annotated sentences (428 from the BIMCV COVID-19 positive partition and 296 from the BIMCV COVID-19 negative partition) referring to COVID-19 as highly likely, uncertain or negative, according to the radiologist's judgment, totaling 23,439 manually annotated sentences. 

\subsubsection*{Annotation of regions of interest}

Since some lesions such as infiltrates, ground glass opacities or consolidation patterns are the most frequent in COVID-19 patients, a sub-set of 10 images were annotated with their ROIs by a team of eight radiologists from the Hospital Universitario de San Juan de Alicante, for the first iteration of the BIMCV COVID-19+ dataset. 

\begin{figure}
    \centering
    \includegraphics[width=\columnwidth]{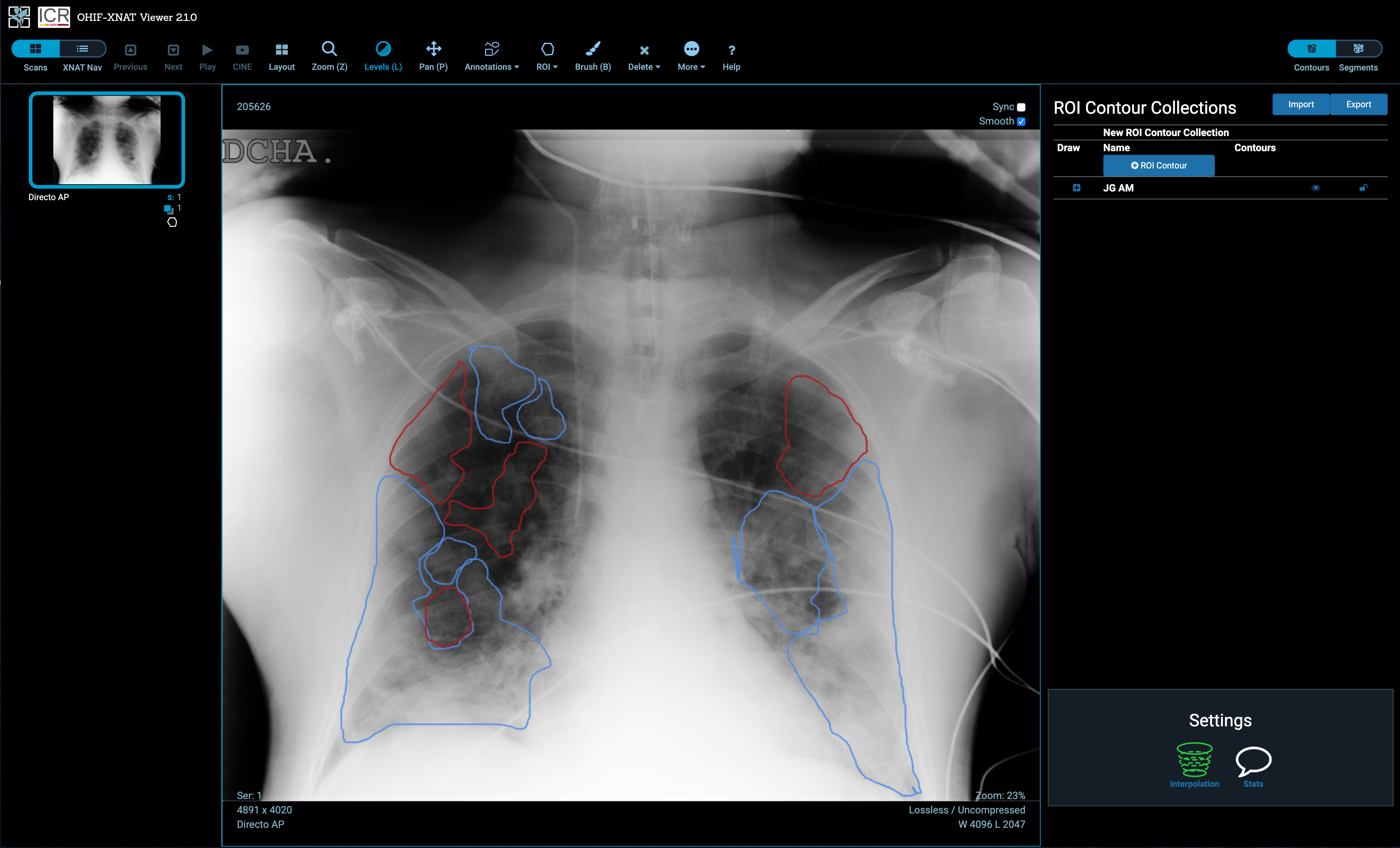}
    \caption{Example annotation at pixel level. In blue, consolidation, red marks ground glass.}
    \label{fig:segmentation}
\end{figure}

As it can be seen in Figure \ref{fig:segmentation}, ROIs corresponding to these findings were labeled at a pixel level using XNAT OHIF Viewer \cite{OHIF}. OHIF is a zero-footprint medical image viewer provided by the Open Health Imaging Foundation that stores an XML output with the ROI paths. This information of great value to train semantic segmentation networks, such as UNet \cite{unet}, in order to extract characteristics such as the extension and exact location of the lesions. Table \ref{tab:findingsSegment} shows the findings annotated at a ROI level.

\begin{table}[!htb]
    \centering
    \begin{tabular}{|l|l|}
    \hline
    Id & Description \\
    \hline
1  & Ground glass XRays                                     \\
2  & Consolidation XRays                                         \\
3  & Peural effusion XRays                                       \\
4  & Interstitial RX                                           \\
5  & Ground glass CT                                     \\
6  & Crazy Paving CT                                           \\
7  & Inverted halo  CT                                         \\
8  & Vascular thickening (in ground glass region) CT \\
9  & Ground glass and consolidation focus CT             \\
10 & Nodular consolidation or in CT            \\
11 & Normal consolidation CT                                 \\
12 & Sprouting tree pattern CT                                       \\
13 & Pleural effusion CT                                        \\
14 & No RX or TAC findings \\ 
\hline
    \end{tabular}
    \caption{Findings annotated with their ROIs using XNAT OHIF Viewer.}
    \label{tab:findingsSegment}
\end{table}


\subsection*{Code availability}

The annotation pipeline from medical reports including their text preprocessing  and the multi-label classifier based on bidirectional Long Short-Term Memory Networks (LSTM) with attention mechanism is available at \url{https://github.com/auriml/Rx-thorax-automatic-captioning}.

\subsubsection*{Database analysis} Statistics regarding the dataset were acquired with a python Jupyter notebook available at \url{https://github.com/BIMCV-CSUSP/BIMCV-COVID-19}.

\section*{Data Records}
\label{sec:datarecords}



The BIMCV specializes in collecting and organizing data imaging information in order to facilitate the research on artificial intelligence and big data with medical imaging. The BIMCV team has designed MIDS (Medical Imaging Data Structure)\footnote{\url{https://github.com/BIMCV-CSUSP/MIDS}}, a standard with which to organize every type of medical information and images in simple hierarchy folders. MIDS constitutes an extension of the BIDS (Brain Imaging Data Structure) standard, a structure that collects medical resonance brain images (MRI) and which is described in \cite{BIDS}.  The objective of MIDS is to expand BIDS, extending its usage to any type of medical imaging modality and anatomical area of the body, rather than just medical brain images.

The main goal of MIDS is to define a standard for the organization and description of medical imaging datasets in the field of artificial intelligence and to facilitate data sharing. For this, we have adapted the DICOM standard in an hierarchical organization that will be based on folder organization with simple file formats (TSV and JSON). The main data types for Imaging are:  nifti (3D) or png (2D) files; for phenotypic data, Tab Separated Value files (.tsv); and for metadata information (Key/value dictionaries), JSON files.

Figure \ref{fig:bids} shows an example of BIMCV COVID-19+ data  structured in MIDS format. The data provided for each sample is described below:

\begin{figure}
    \centering
    \includegraphics[width=\columnwidth]{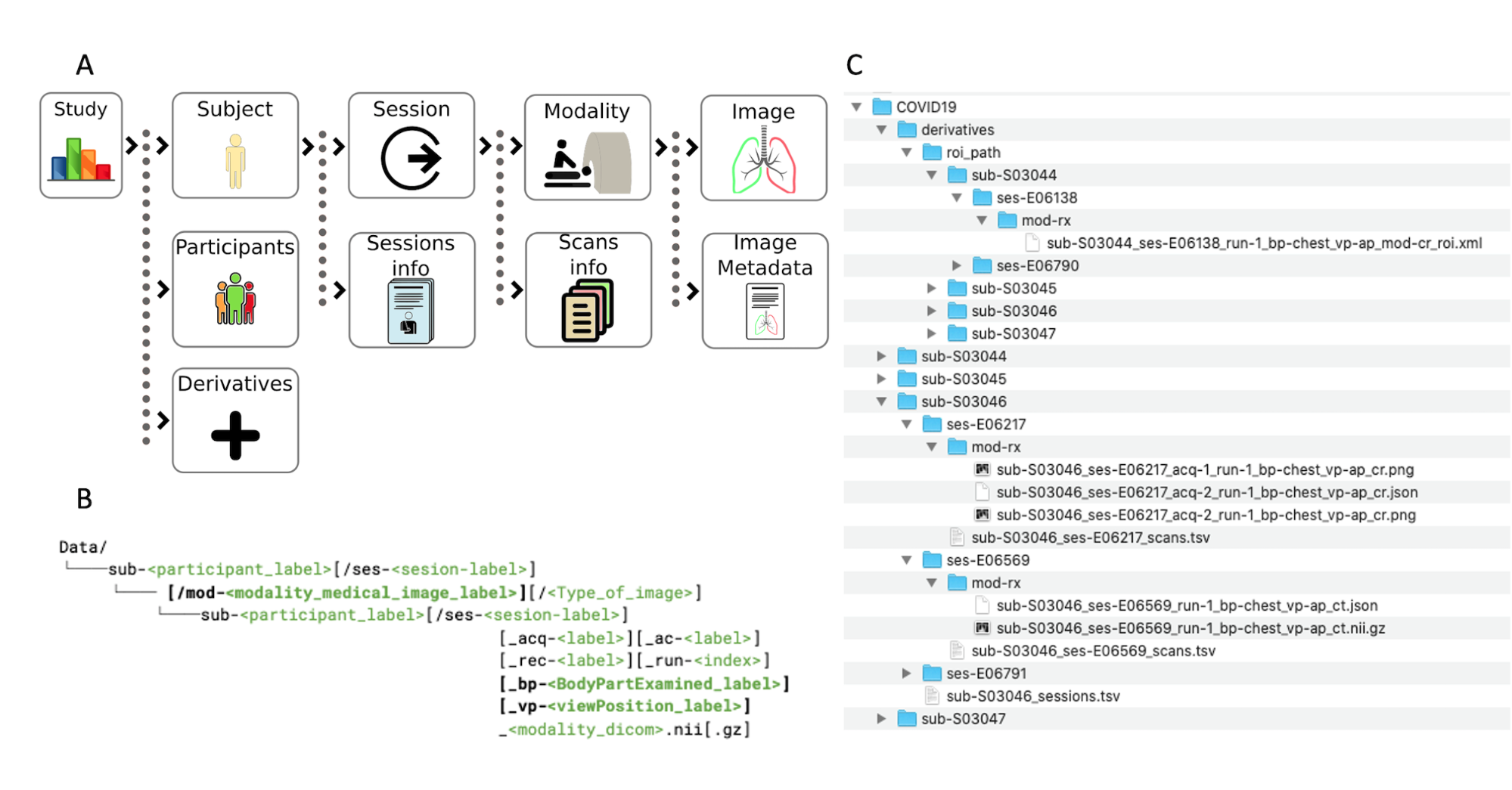}
    \caption{MIDS structure for BIMCV COVID-19+ dataset. (A) Conceptual schema (B) General Template (C) Example of folder structure.}
    \label{fig:bids}
\end{figure}

\textbf{Diagnostic tests:} All the diagnostic tests performed on each individual to diagnose COVID-19. The diagnostic test could be PCR, IGG or IGM. The result of the test can be positive, negative or indeterminate. A subject can have several tests during the period of time with different outcomes. The list of tests is stored in the file \textit{sil\_reg\_covid.csv}.

\textbf{Radiological reports:} The radiological reports for each image study performed on the subject anonymized, as explained in radiological reports anonymization section. They are included in the file\\
\textit{derivatives/EHR/labels/labels\_COVID-19\_posi.tsv}.

\textbf{CUI disease and location terms: } extracted automatically using the methodology described in  Labeling subsection. They are stored in the same file as the radiological reports.

\textbf{CUI hierarchy: } a tree description of the CUI hierarchy used to describe the diseases. File:\\ \textit{/derivatives/EHR/labels/tree\_term\_CUI\_counts\_image\_covid\_posi.csv}.

\textbf{Image data: } Each subject can have a plurality of image studies, also called sessions, performed during the period of time of the observation. Each image study has a plurality of image series. They are all stored in the directory structure depicted in Fig.~\ref{fig:bids}. Image data is extracted from the DICOM images and stored in \textit{.nii.gz}. The relevant DICOM fields for each image series are stored in a JSON file with the same name as the image series.



\section*{Technical Validation}

\begin{figure}[!ht]
    \centering
    \includegraphics[width=0.6\columnwidth]{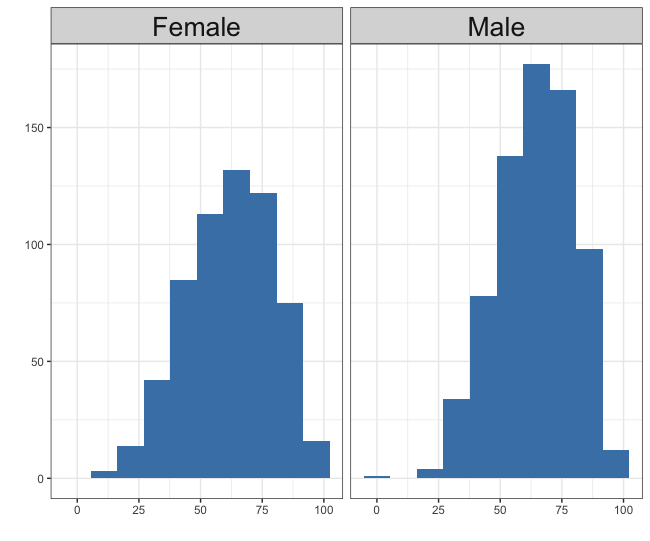}
    \includegraphics[width=0.6\columnwidth]{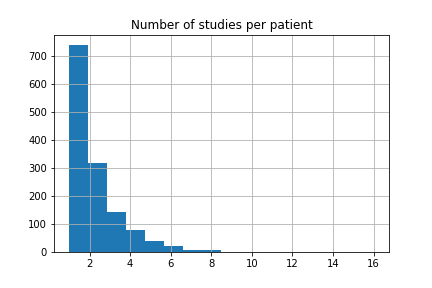}
    \includegraphics[width=0.6\columnwidth]{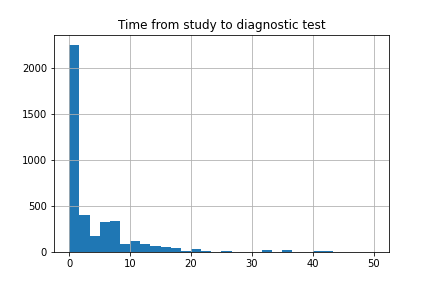}
    \caption{Top: histogram of the patients age. Middle: histogram of the number of studies per subject. Bottom: histogram of the difference (in days) between the image study and the diagnostic test. Please note that for most of the studies, there were less than five days between the radiography and the diagnostic test.}
    \label{fig:hists}
\end{figure}

The data collected contains 1,311 subjects, 2,429 image studies and 5,530 image series. 602 patients were female (45.92\%). The mean age was of 63.11 ($\pm$ 16.75). The histogram of the patients' age is shown in Fig.~\ref{fig:hists} (top) and is highly coherent with the demographics of COVID-19+ in Spain. 

There is an average of 1.9 image studies per subject. The distribution can be seen in Fig.~\ref{fig:hists} (middle). With respect to image modality, the dataset contains 1,380 CX, 885 DX and 163 CT studies. 2,427 chest x-rays were acquired in monochrome 2 photometric interpretation and 751 in monochrome 1. Images in monochrome 1 should be inverted to be visualized correctly. The vendors of the devices used to acquire the studies are shown in Table~\ref{tab:machines}. As can be seen, there are different acquisition systems, which ensures variability in the dataset images.

\begin{table}[h!]
\centering
\begin{tabular}{|l|l|}
\hline
 Manufacturer and model & \#studies\\
\hline \hline
KONICA MINOLTA 0862                        &       342 \\
GMM ACCORD DR                              &       255 \\
SIEMENS SIEMENS FD-X                       &       247 \\
Agfa DR 14e C - 1200ms                     &       161 \\
Agfa DX-M                                  &       161 \\
"GE Healthcare" "Thunder Platform"         &       150 \\
Philips Medical Systems DigitalDiagnost    &       133 \\
Philips Medical Systems PCR Eleva          &       113 \\
Agfa 3543EZE                               &       112 \\
Carestream Health DRX-1                    &        87 \\
KONICA MINOLTA CS-7                        &        77 \\
SIEMENS SOMATOM go.Up                      &        69 \\
Agfa Pixium\_4343E\_CSI                      &        59 \\
Agfa CR30-X                                &        39 \\
Canon Inc. CXDI Control Software NE        &        37 \\
TOSHIBA Aquilion                           &        36 \\
Philips DigitalDiagnost                    &        34 \\
KONICA MINOLTA 0110                        &        23 \\
Philips Brilliance 16                      &        23 \\
Philips Medical Systems Essenta DR         &        21 \\
\hline
\end{tabular}
\caption{\label{tab:machines} Number of studies acquired by each device for devices with more than 15 studies acquired.}
\end{table}

A total of 2,425 PCR tests have been performed on the patients, of which 1,773 were positive to SARS-Cov-2, 622 negative to SARS-Cov-2 and 30 indeterminate; 29 IGM, 29 IGG and 35 ACT tests were performed, of which 44 were positive, 32 negative and 17 indeterminate. The average closest time between an image and a diagnostic test is 5.03 days, the distribution being that shown in Fig.~\ref{fig:hists} (bottom). Please note that some images have a diagnostic test as many as 50 days afterwards. This is acceptable, since the period of observation is two months and most image studies were  performed on the subjects prior to their COVID-19 diagnosis.

135 studies were deemed suboptimal in the radiology reports. The most common radiological findings for the images, along with their number of appearances, are shown in Table~\ref{tab:diagnoses}. Owing to the construction of the database, COVID-19 appears as the most common radiological finding, followed by densities, pneumonia, consolidations and infiltrates, all closely related to COVID-19.

Image anonymization was validated by visually inspecting all the cropped images to verify that none of the patients' data were displayed.

Chest x-ray images were manually inspected to validate whether the patient orientation was correctly estimated with the neural network described in the Image Preprocessing section. The orientation was changed if it was mistakenly estimated.

\begin{table}[h!]
\centering
\begin{tabular}{|l|l|}
\hline
Radiological finding and diagnosis &    Number \\
\hline \hline
COVID 19                                    &  802 \\
increased density                           &  657 \\
pneumonia                                   &  605 \\
unchanged                                   &  494 \\
consolidation                               &  357 \\
infiltrates                                 &  346 \\
interstitial pattern                        &  317 \\
alveolar pattern                            &  291 \\
normal                                      &  284 \\
ground glass pattern                        &  240 \\
cardiomegaly                                &  166 \\
pleural effusion                            &  149 \\
laminar atelectasis                         &   88 \\
costophrenic angle blunting                 &   83 \\
suboptimal study                            &   82 \\
viral pneumonia                             &   62 \\
endotracheal tube                           &   59 \\
aortic elongation                           &   58 \\
nodule                                      &   54 \\
central venous catheter                     &   54 \\
\hline
\end{tabular}

\caption{\label{tab:diagnoses} Radiological findings and diagnoses extracted from the radiology reports when having more than 50 appearances.}
\end{table}





Regarding the labeling of the reports, in some of those in which the text ``COVID-19'' was mentioned, those mentions were actually negated expressions for COVID-19 disease based on the image. The neural network was trained to ensure that the model could detect these negative COVID-19 cases and validate that the labels ``COVID-19'' and ``COVID-19 uncertain'' were correctly extracted. The model achieved an $F1$-micro of 0.922 for the validation set, with 2,343 manually labeled sentences (including sentences from Padchest \cite{padchest}, BIMCV COVID-19 positive and BIMCV COVID-19 negative partitions), and was further validated using an independent test sample. 

This independent test sample consisted of 64 new sentences manually labeled and randomly extracted from the BIMCV COVID-19 negative partition in which the tern COVID-19 was mentioned. In some cases, it was mentioned negatively (24), in others it was denoted as low suspicion based on images with atypical findings ``COVID 19 uncertain'' (13), and in the rest (27) it was mentioned affirmatively because the image was compatible with COVID-19 even if those studies were included in the BIMCV-COVID-19 negative samples. Results for this experiment are shown in Table \ref{tab:results}.

\begin{table}[h!]
\centering
\begin{tabular}{|l|l|l|l|}
\hline
Radiological findings set &
Precision&
Recall&
F-Score\\
 \hline \hline
COVID-19&
0.961&
0.925&
0.943\\
 \hline
COVID-19 uncertain&
1.0&
0.846&
0.916\\
 \hline

\end{tabular}

\caption{\label{tab:results} Classification results for an independent sample test of the COVID-19 labels}
\end{table}

In addition, the method was evaluated for all included entities, both those related or not related to COVID-19 pneumonia, in the same independent test set, using the metrics described in \cite{padchest}, obtaining an $F1$-weighted=$0.9320$, $F1$-micro=$0.9378$, and accuracy=$0.8281$.

\begin{figure}
    \centering
    \includegraphics[width=\columnwidth]{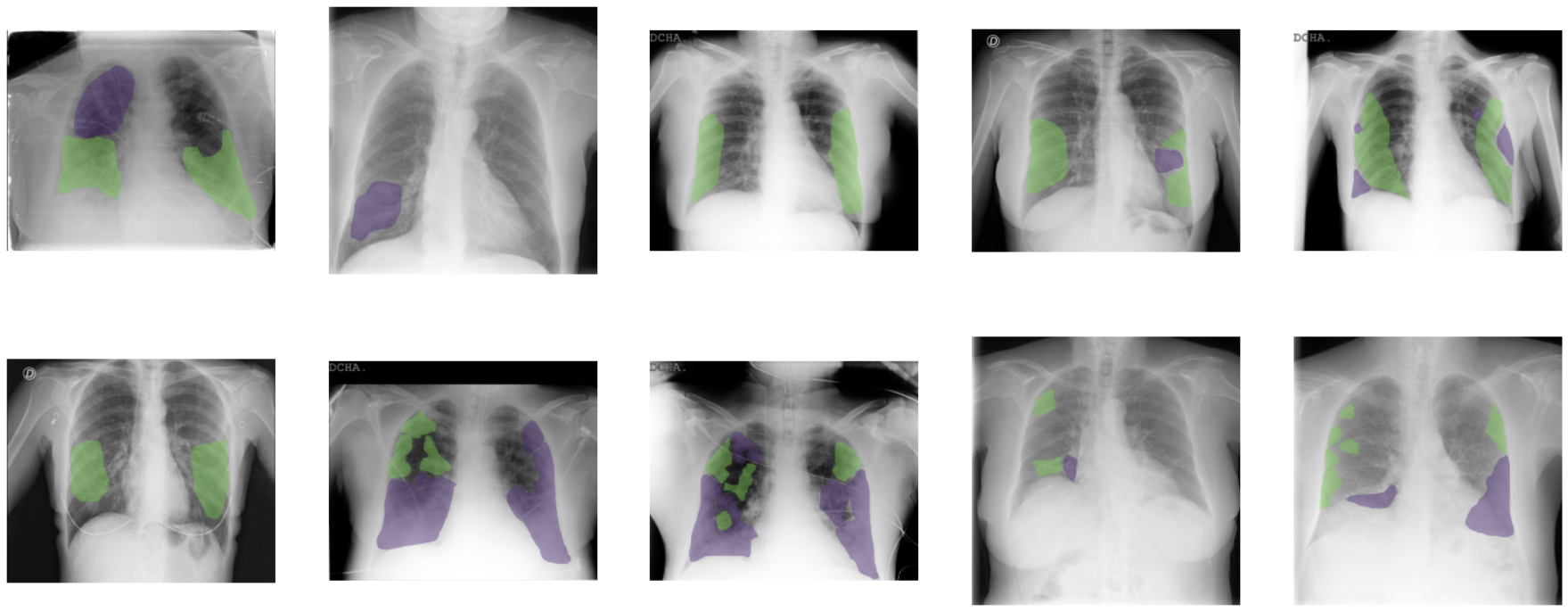}
    \caption{ROIs marked for 10 images from different series. Green: ground glass opacities. Purple: consolidations.}
    \label{fig:rois}
\end{figure}

\begin{figure}
    \centering
    \includegraphics[width=\columnwidth]{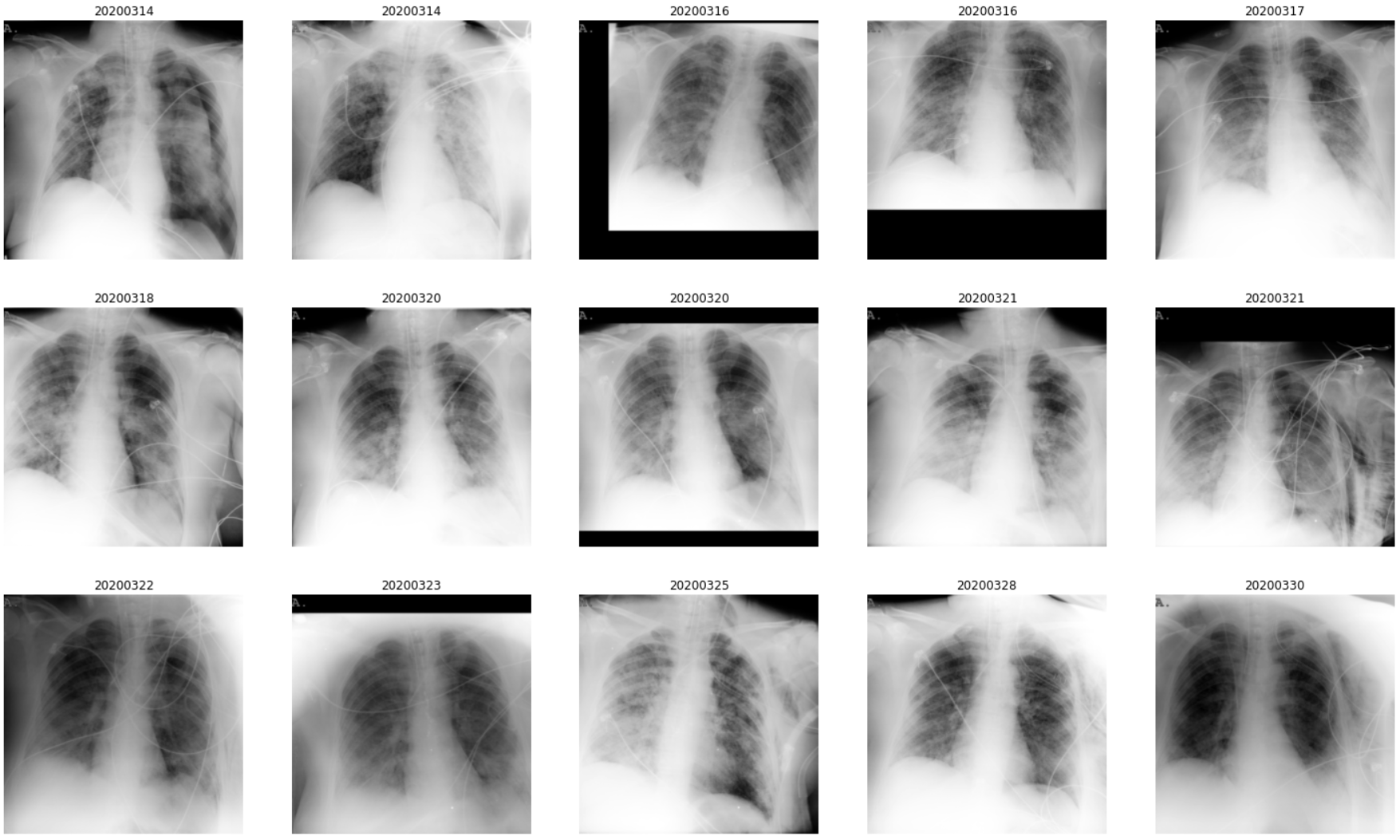}
    \caption{Longitudinal series for the same subject with 15 image studies between March 14th and March 30th 2020.  }
    \label{fig:longserie}
\end{figure}

\section*{Usage Notes}

This dataset provides valuable information to professionals and could, for example, be used to train radiologists in this new pathology and understand the lesions that can be typically found. It additionally serves to train deep learning methods with the images and labels in order to assist radiologists in the decision-making process. 

The data can be downloaded on request from the BIMCV web page (\url{http://bimcv.cipf.es/bimcv-projects/bimcv-covid19}). It is freely available for research, and can also be used for commercial purposes under certain conditions. Before downloading the data, the user should accept the End-User License Agreement detailed in \url{http://bimcv.cipf.es/bimcv-projects/bimcv-covid19/bimcv-covid19-dataset-research-use-agreement-2/}.

The dataset is located in two repositories: 1) the center for Open Science (\url{https://osf.io}) in Germany, with DOI  10.17605/OSF.IO/NH7G8, and 2) EUDAT, the largest infrastructure of integrated data services and resources supporting research in Europe through TransBioNet - Task Force COVID-19 in Barcelona Supercomputing Center (BSC).


\section*{Acknowledgements}

This work is first and foremost an open and free contribution from the authors in the working group with support from the Regional Ministry of Innovation, Universities, Science and Digital Society grant awarded through decree 51/2020 by the Valencian Innovation Agency (Spain) and Regional Ministry of Health in Valencia Region. This research is also supported by the University of Alicante's UACOVID-19-18 project. 

Part of the infrastructure used has been cofunded by the European Union through the Operational Program of the European Fund of Regional Development (FEDER) of the Valencian Community 2014-2020. The Medical Image Bank of the Valencian Community was partially funded by the European Union's Horizon 2020 Framework Programme under grant agreement 688945 (Euro-BioImaging PrepPhase II).

We are grateful to General Electric Healthcare, which altruistically made available the technical and human means necessary to perform parameterized searches, and to the Bioinformatics and Biostatistics Unit of the CIPF, which is located within the ‘TransBioNet - Task Force COVID-19 Processing and analyzing samples’, for sharing the cluster network, thus facilitating the downloads. We are also grateful to NVIDIA for the generous donation of a Titan Xp and a Quadro P6000.

\section*{Author contributions}

Study design: M.I.V., J.M.S., A.B.,  Data curation: M.I.V., J.M.S., J.A.M., J.M.S., A.B., M.A.C., J.G., G.G.,  Technical validation: A.B., G.G., A.P., X.B., F.G.-G., Oversaw project: M.I.V, D.O.-B., A.P., Manage field operations: J.M.S. All authors contributed to writing and editing the manuscript.




\section*{Competing interests}

The authors declare no competing interests.

\bibliographystyle{naturemag}
\bibliography{refs}

\end{document}